\documentstyle[epsf]{aipproc}

\begin{document}
\title{Large-$N$ Yang--Mills Theory as Classical Mechanics}

\author{C.-W. H. Lee\footnote{speaker.} and S. G. Rajeev}
\address{Department of Physics and Astronomy, University of Rochester, P.O. Box 270171, Rochester, New York 14627.}

\maketitle

\begin{abstract}
To formulate two-dimensional Yang--Mills theory with adjoint matter fields in the large-$N$ limit as classical 
mechanics, we derive a Poisson algebra for the color-invariant observables involving adjoint matter fields.  We
showed rigorously in Ref.\cite{lera} that different quantum orderings of the observables produce essentially the 
same Poisson algebra.  Here we explain, in a less precise but more pedagogical manner, the crucial topological 
graphical observations underlying the formal proof. 
\end{abstract}

One major unsolved problem in physics is hadronic structure.  We would like to explain, for instance, the momentum 
distributions of valence quarks, sea quarks and gluons inside a proton.  We have accumulated a fairly large amount 
of experimental data on the distribution functions \cite{cteq}, but we have made relatively little advance in 
explaining them from the widely accepted fundamental theory of strong interaction, quantum chromodynamics (QCD).

The emergence of hadrons is a low-energy phenomenon of strong interaction.  The strong coupling constant is large, 
and perturbative QCD fails very badly.  Other approximations are needed.  One widely studied approximation is the
large-$N$ limit, in which the number of colors $N$ is taken to be infinitely large \cite{thooft74a}.  This 
approximation is believed to capture the essence of low-energy strong interaction phenomena.  Indeed, 't Hooft 
showed that if space--time is assumed to be two-dimensional only, then the meson spectrum displays itself as a Regge
trajectory \cite{thooft74b}.  This attractive feature of the large-$N$ limit has drawn the attention of a large 
number of researchers for more than two decades.  They want to build up a systematic theory of the large-$N$ limit 
to deal with baryons in addition to mesons in four-dimensional space--time.

One important feature of the large-$N$ limit is that the expectation value of a product of two observables $A$ and 
$B$ is the same as the product of the expectation values of these two observables.  The difference is of order 
$1/N$ and so can be omitted \cite{witten,coleman}:
\[ \langle \hat{A} \hat{B} \rangle = \langle \hat{A} \rangle \langle \hat{B} \rangle + O(1/N). \]
In other words, there is no quantum fluctuation.  The theory thus behaves like {\em classical mechanics} 
\cite{yaffe}, and it should be possible for us to formulate the large-$N$ limit of QCD as classical mechanics.

Formally speaking, we need three ingredients to build up a theory as classical mechanics \cite{arnold}.  The first 
is the notion of a manifold to describe the geometry of the phase space of positions and momenta.  Dynamical 
variables are then functions on the manifold.  The second is the notion of a Hamiltonian function $H$, one of the 
dynamical variables.  This function displays the physical features (e.g., symmetry) of the system, and governs the 
time evolution of it.  How the Hamiltonian function governs the time evolution is determined by a {\em Poisson 
algebra} \cite{chpr}, the third ingredient of classical mechanics.  To understand what a Poisson algebra is, we 
need a number of preliminary notions.

An {\em algebra} $R$ is a linear space on a field $K$ with a multiplication rule of any 2 vectors in $R$ such that 
for any $x$, $y$ and $z \in R$ and $a \in K$, 
\begin{enumerate}
\item $xy \in R$;
\item the vectors satisfy the distributive properties 
\begin{eqnarray*}
   x(y + z) & = & xy + xz \; \mbox{and} \\ 
   (x + y)z & = & xz + yz \mbox{; and} 
\end{eqnarray*}
\item $a (xy) = (ax) y = x (ay)$.
\end{enumerate}
An {\em associative algebra} is an algebra $R$ such that for any $x$, $y$ and $z \in R$, 
\[ x (yz) = (xy) z. \]
A {\em Poisson algebra} is an associative algebra $R$ which is equipped with a blinear map $\{ \; , \; \}: R \times 
R \rightarrow R$, the Poisson bracket, with the following properties for any $x$, $y$ and $z \in R$:
\begin{enumerate}
\item skew-symmetry: 
\[ \{ y, x \} = - \{ x, y \} ; \]
\item the Jacobi identity: 
\[ \{ x, \{ y, z \} \} + \{ y, \{ z, x \} \} + \{ z, \{ x, y \} \} = 0 \mbox{; and} \]
\item the Leibniz identity: 
\[ \{ xy, z \} = x \{ y, z \} + \{ x, z \} y. \]
\end{enumerate}
How a dynamical variable $G(t)$ changes with the time $t$ is given by the equation
\[ \frac{dG(t)}{dt} = \{ G. H \}. \]

Making the assumption that space--time is two-dimensional, one of us introduced a formulation of large-$N$
Yang--Mills theory as classical mechanics several years ago \cite{rajeev}.  As in the mesonic model 't Hooft 
studied, gluons are not dynamical objects in this classcial mechanics; only quarks and anti-quarks are.  The phase 
space turns out to an infinite-dimensional Grassmannian manifold \cite{prse}.  (Briefly speaking, an 
infinite-dimensional Grassmannian manifold is a collection of subspaces of an infinite-dimensional vector space.  A 
differential structure is conferred upon this collection to turn it into a manifold.)  Dynamical variables are 
composed of quark and anti-quark fields.  They are bilocal functions on the Grassmannian manifold.  The Poisson 
bracket can be uniquely determined from the geometric properties of the Grassmanian.  The Hamiltonian is chosen in 
such a way that upon quantization of this classical mechanics, i.e., if we retain terms of subleading orders in the 
large-$N$ limit, it will take exactly the form the action of Yang--Mills theory is conventionally written.  The 
meson spectrum predicted by this classical mechanics is precisely the same as that obtained by 't Hooft.  However, 
the model can be used to calculate structure functions of a baryon also \cite{krra}.  As an initial attempt, the 
proton is assumed to be made up of valence quarks only.  Sea quarks and gluons are omitted.  In the infinite 
momentum frame, the transverse momenta of valence quarks can be neglected.  Thus the Yang--Mills theory can be 
dimensionally reduced from four dimensions to two.  Dimensional reduction is a good approximation for valence 
quarks carrying a large fraction of the total momentum of the proton, but not so good for the quarks carrying 
almost no momentum.  Indeed, it turned out the momentum distribution functions predicted by this classical model of 
Yang--Mills theory agrees well with experimental data at high-momentum regime, but not so well near zero momentum.  

How about sea quarks and gluons inside a hadron?  The sea quark distribution function can actually be predicted 
within the same classical mechanical framework \cite{rajeev,jkr}; it is a matter of modifying the details of some
approximations.  However, gluon dynamics require a new set Poisson brackets and a new phase space.  Since gluons
carry about 20\% of the total momentum of a proton \cite{cteq}, it is worthwhile to study them further.

\begin{figure}[ht]
\epsfxsize 5in
\centerline{\epsfbox{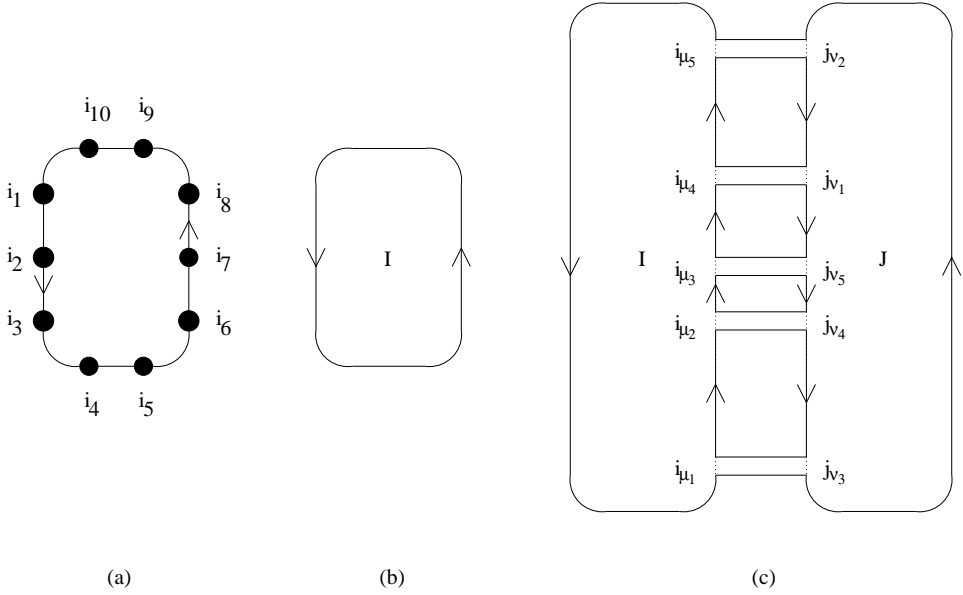}}
\caption{(a) A typical color-invariant observable $f^I$.  Each solid circle represents an $\eta^i$.  Notice the 
cyclic symmetry of the figure.  (b) A simplified diagrammatic representation of $f^I$.  We use the capital letter 
$I$ to denote the whole sequence $i_1$, $i_2$, \ldots, $i_m$.  (c) A typical term in $\{ f^I, f^J \}_W$.  This is a 
product of the  color-invariant observables $f^{I(\mu_1, \mu_2) J(\nu_4, \nu_3)}$, $f^{I(\mu_2, \mu_3) 
J(\nu_5, \nu_4)}$, $f^{I(\mu_3, \mu_4) J(\nu_1, \nu_5)}$, $f^{I(\mu_4, \mu_5) J(\nu_2, \nu_1)}$, and 
$f^{I(\mu_5, \mu_1) J(\nu_3, \nu_2)}$.  We can identify these color-invariant observables by their vertices.  For
example, $f^{I(\mu_1, \mu_2) J(\nu_4, \nu_3)}$ can be described as a `loop with vertices $i_{\mu_1}$, $i_{\mu_2}$,
$j_{\nu_4}$ and $j_{\nu_3}$', though none of these vertices belong to $f^{I(\mu_1, \mu_2) J(\nu_4, \nu_3)}$.}
\label{f1}
\end{figure}

A Poisson algebra for gluons was constructed a few years ago \cite{ratu}.  This was achieved by a technique called
deformation quantization \cite{fllist,chpr}.  Deformation quantization refers to the procedure of defining an
algebra of smooth functions in such a way that when the functions are multiplied, it is as if we are multiplying 
suitably ordered operators these smooth functions represent.  Any physical observable of gluons must involve 
creation and annihilation operators of gluons, each of which carries two color quantum numbers and a vector-valued 
linear momentum.  The two color quantum numbers can be treated as the row and column indices of an $N \times N$ 
matrix, $N$ being the total number of colors.  Any physical observable has to be color-invariant.  This implies 
that it has to be a polynomial of the traces of matrix products with a generic form  
\[ f^I = \frac{1}{N^{m/2 + 1}} {\rm Tr} \eta^{i_1} \eta^{i_2} \cdots \eta^{i_m}, \]
where $i_1$, $i_2$, \ldots, $i_m$ are quantum states other than colors of the gluons, $I$ is the sequence $i_1$, 
$i_2$, \ldots $i_m$, and the factor of $N$ is put to the left of the trace to ensure that the Poisson algebra of 
these $f^I$'s, to be introduced shortly, is well defined in the large-$N$ limit.  In each product, the operators 
have to satisfy a certain ordering.  As an initial attempt, the operators are Weyl-ordered in Ref.\cite{ratu}.  
Hence,
\begin{eqnarray*}
   \eta^i & = & 1/2 (a^{i\rho}_{\sigma} + a^{\dagger i \rho}_{\sigma}) \\
   \eta^{-i} & = & i/2 (a^{i\rho}_{\sigma} - a^{\dagger i \rho}_{\sigma})
\end{eqnarray*}
When we multiply two physical observables together, we need to rearrange the order of the creation and annihilation 
operators to make the resultant product consistent with the quantum ordering.  As a result, multiplication of 
physical observables is still associative but no longer commutative.  The commutator of two physical observables 
therefore provides us a Poisson bracket.  In the case of Weyl-ordered quantum observables of gluons, the Poisson
bracket is
\begin{eqnarray}
   \{ f^I, f^J \}_W & = & 2i \sum_{r=1, \mbox{odd}}^{\infty}
   \sum_{\begin{array}{l}
            \mu_1 < \mu_2 < \cdots < \mu_r \\
            (\nu_1 > \nu_2 > \cdots > \nu_r) 
         \end{array}}
   (-\frac{i\hbar}{2})^r 
   \tilde{\omega}^{i_{\mu_1} j_{\nu_1}} \cdots \tilde{\omega}^{i_{\mu_r} j_{\nu_r}} \nonumber \\
   & & \cdot f^{I(\mu_1, \mu_2)J(\nu_2, \nu_1)} f^{I(\mu_2, \mu_3)J(\nu_3, \nu_2)}
   \cdots f^{I(\mu_r, \mu_1)J(\nu_1, \nu_r)}.
\label{1}
\end{eqnarray}
In this equation, $\tilde{\omega}^{ij}$ is an anti-symmetric constant tensor.  $I(\mu_1, \mu_2) J(\nu_2, \nu_1)$ is
the sequence $i_{\mu_1 + 1}$, $i_{\mu_1 + 2}$, \ldots, $i_{\mu_2 - 1}$, $j_{\nu_2 + 1}$, $j_{\nu_2 + 2}$, \ldots,
$j_{\nu_1 - 1}$.  Notice that we sum over all possible sets of values of $\mu_1$, $\mu_2$, \ldots, $\mu_r$ such that
they are strictly increasing, and all possible sets of values of $\nu_1$, $\nu_2$, \ldots, $\nu_r$ such that they 
are strictly decreasing up to a cyclic permutation.  Eq.(\ref{1}) looks complicated, but we can actually visualize 
it in Fig.~\ref{f1}.  We will call this Poisson algebra ${\cal W}$.

Nevertheless, in many practical applications, we need quantum observables made up of normal-ordered rather than 
Weyl-ordered operators.  We thus need to apply the above idea of deformation quantization to derive a Poisson 
bracket of normal-ordered observables \cite{lera}.  In this case, A color-invariant observable has the form
\[ \phi^I = \frac{1}{N^{n/2 + 1}} {\rm Tr} z^{i_1} z^{i_2} \cdots z^{i_n}, \]
where 
\[ z^{i\rho}_{\sigma} = a^{i\rho}_{\sigma} \; \mbox{and} \; z^{-i\rho}_{\sigma} = a^{\dagger i \rho}_{\sigma} \]
for $i > 0$.  The Poisson bracket for these operators turns out to be
\begin{eqnarray}
   \{\phi^{I},\phi^{J}\}_N & = & \sum_{r=1}^{\infty}
   \sum_{\begin{array}{l}
     \mu_{1}<\mu_{2}<\ldots <\mu_{r} \\
     (\nu_{1}>\nu_{2}>\ldots >\nu_{r})
   \end{array}}
    \hbar^{r} \gamma^{i_{\mu_{1}} j_{\nu_{1}}} \ldots
    \gamma^{i_{\mu_{r}} j_{\nu_{r}}} \nonumber \\
    & & \cdot \phi^{I(\mu_{1},\mu_{2})J(\nu_{2},\nu_{1})}
    \phi^{I(\mu_{2},\mu_{3})J(\nu_{3},\nu_{2})} \ldots
    \phi^{I(\mu_{r},\mu_{1})J(\nu_{1},\nu_{r})} - (I\leftrightarrow J),
\label{2}
\end{eqnarray}
where $\gamma^{\mu}_{\nu} = 0$ unless $\mu < 0$ and $\nu > 0$.  We will call this Poisson algebra ${\cal N}$.

The Poisson algebras ${\cal W}$ and ${\cal N}$ look very different.  Are they intrinsically different Poisson 
algebras, or are they the same Poisson algebra with different expressions?  This question can be answered by 
looking for a Poisson morphism between ${\cal W}$ and ${\cal N}$.  Let $R_1$ and $R_2$ be two Poisson algebras.  A 
{\em Poisson morphism} is a mapping $F: R_1 \rightarrow R_2$ such that it preserves
\begin{enumerate}
\item vector addition: 
\[ F(x + y) = F(x) + F(y); \]
\item scalar multiplication: 
\[ F(kx) = k F(x); \]
\item vector multiplication: 
\[ F(xy) = F(x) F(y) \mbox{; and} \] 
\item the Poisson bracket: 
\[ F( \{ x, y \}_1 ) = \{ F(x), F(y) \}_2 \]
\end{enumerate}
for any $k \in K$, $x$ and $y \in R_1$.  Here $\{ \; , \; \}_1$ and $\{ \; , \; \}_2$ are the Poisson brackets of 
$R_1$ and $R_2$, respectively.  If there exists a Poisson morphism between two Poisson algebras, then they are 
effectively the same Poisson algebra. 

\begin{figure}
\epsfxsize 5.5in
\centerline{\epsfbox{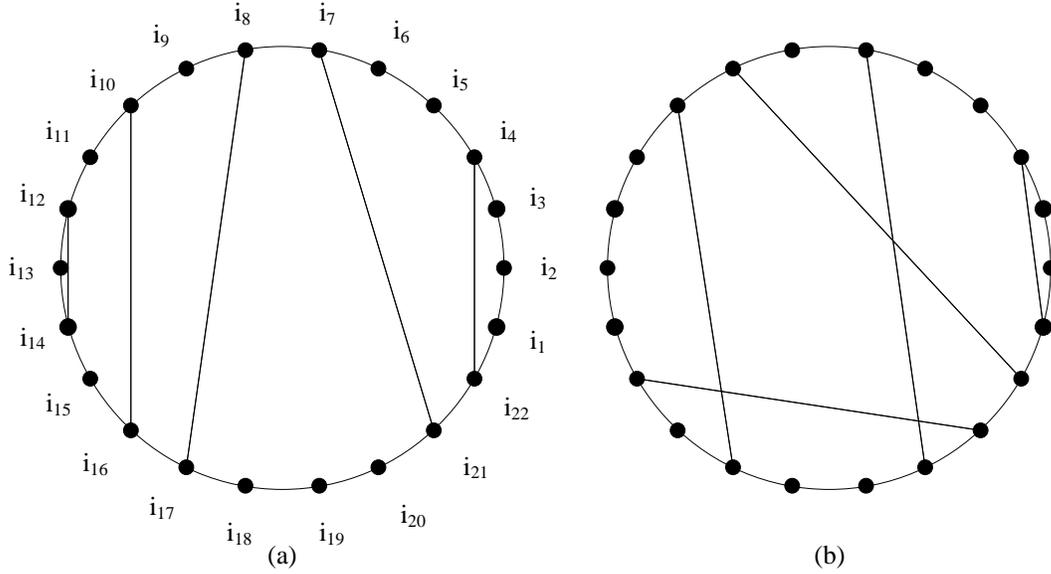}}
\caption{(a) An allowable partition of a color-invariant observable into a product of color-invariant observables.
Note that no two lines cross each other.  (b) A forbidden partition of the color-invariant observable.  Note that
some lines cross one another.}
\label{f2}
\end{figure}

In our case, a Poisson morphism $F: {\cal W} \rightarrow {\cal N}$ {\em does exist}.  The reader can find a 
thoroughly rigorous formulation and proof of this Poisson morphism in Ref.\cite{lera}.  Roughly speaking, the 
mapping $F$ is accomplished in two steps.  Consider the color-invariant observable $f^{i_1 \ldots i_{22}} \in 
{\cal W}$ in Fig.~\ref{f2}.  The first step involves splitting the big all-encompassing loop into a number of 
smaller loops.  The straight lines joining the solid circles and cutting off the big loop have to be within the 
loop.  Moreover, no two straight lines can cross each other.  Each solid circle $\eta^i$ which is not touched by 
any straight line is now identified as a linear combination of solid circles $\eta'^i$ in ${\cal N}$ by the formula
\[ \eta'^i = \left\{ \begin{array}{ll} \frac{1}{2} (z^i + z^{-i}) & \mbox{if $i > 0$; and} \\
             \frac{\rm i}{2} (z^i - z^{-i}) & \mbox{if $i < 0$.} \end{array} \right. \]
The resultant diagram represents the product of the color-invariant observables in ${\cal N}$, each of which is 
represented by a smaller loop in the resultant diagram.  For instance, the product in Fig.~\ref{f2}(a) is
\[ T^{i_4 i_{22}} T^{i_7 i_{21}} T^{i_8 i_{17}} T^{i_{10} i_{16}} T^{i_{12} i_{14}} \\
   \phi^{i_1 i_2 i_3} \phi^{i_5 i_6} \phi^{i_9} \phi^{i_{11} i_{15}} \phi^{i_{13}} \phi^{i_{18} i_{19} i_{20}}, \]
where $T^{ij}$ is a polynomial of $C^{ij}$ and $C^{ji}$, where in turn $C^{ij}$ is a polynomial of $\gamma^{ij}$, 
$\gamma^{-i, j}$, $\gamma^{i, -j}$ and $\gamma^{-i, -j}$ \cite{lera}.  The second step involves summing over all 
allowable partitions of $f^I$ in ${\cal W}$.

\begin{figure}
\epsfxsize 5.1in
\epsfysize 4.1in
\centerline{\epsfbox{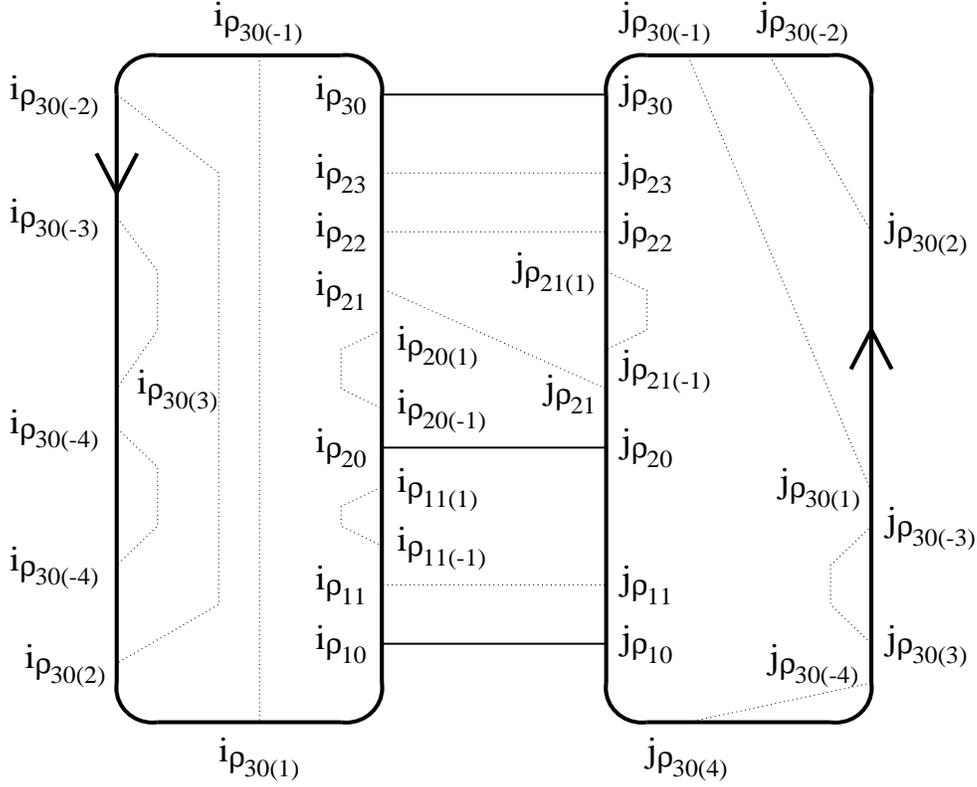}}
\caption{A typical term in $\{ F(f^I), F(f^J) \}_N$ or $F ( \{ f^I, f^J \}_W )$.}
\label{f3}
\end{figure}

Why is this $F$ a Poisson morphism?  The most non-trivial statement we need to show is that $F$ preserves the 
Poisson bracket.  In other words, We need to show that each term in $\{ F(f^I), F(f^J) \}_N$ is a term in
$F( \{ f^I, f^J \}_W )$, and vice versa.  Consider Fig.~\ref{f3}.  The big oval-shaped object on the left is $f^I$, 
and the one on the right is $f^J$.  The dotted lines inside $f^I$ divide it into a product of color-invariant 
observables in ${\cal N}$.  So do the dotted lines insider $f^J$.  A typical term in $\{ F(f^I), F(f^J) \}_N$ is
obtained by choosing a smaller loop in $I$, and a smaller loop in $J$, and perform the contractions illustrated in
Fig.~\ref{f1}.  In Fig.~\ref{f3}, we have chosen the smaller loop with vertices $i_{\rho_{11(-1)}}$, 
$i_{\rho_{11(1)}}$, $i_{\rho_{20(-1)}}$, $i_{\rho_{20(1)}}$, $i_{\rho_{30(-1)}}$ and $i_{\rho_{30(1)}}$ in $I$, and
the smaller loop $j_{\rho_{21(-1)}}$, $j_{\rho_{21(1)}}$, $j_{\rho_{30(-1)}}$, $j_{\rho_{30(1)}}$, 
$j_{\rho_{30(-3)}}$, $j_{\rho_{30(3)}}$, $j_{\rho_{30(-4)}}$ and $j_{\rho_{30(4)}}$ in $J$.  (See the explanation
of the jargons here in the caption of Fig.~\ref{f1}.)  There are 7 contractions due to the Poisson bracket.  The 
contracted pairs are $i_{\rho_{10}}$ and $j_{\rho_{10}}$, $i_{\rho_{11}}$ and $j_{\rho_{11}}$, $i_{\rho_{20}}$ and 
$j_{\rho_{20}}$, $i_{\rho_{21}}$ and $j_{\rho_{21}}$, $i_{\rho_{22}}$ and $j_{\rho_{22}}$, $i_{\rho_{23}}$ and 
$j_{\rho_{23}}$, and $i_{\rho_{30}}$ and $j_{\rho_{30}}$.  According to Eq.(\ref{1}), these contractions produce in 
this term of $\{ F(f^I), F(f^J) \}_N$ a constant factor $G$ which the reader can divine is a polynomial of 
$C^{\pm i_{\rho_{10}}, \pm j_{\rho_{10}}}$, \ldots, $C^{\pm i_{\rho_{30}}, \pm j_{\rho_{30}}}$.  Now notice a 
crucial observation.  Identify $\tilde{\omega}^{ij}$ with a certain polynomial of $C^{ij}$ and $C^{ji}$.  Replace 
an {\em odd} number of $C^{ij}$'s in $G$ with $\tilde{\omega}^{ij}$'s, and the remaining $C^{ij}$'s with 
$T^{ij}$'s.  Add up all such possible replacements.  The sum will precisely be $G$ up to a multiplicative constant.
Consequently, we can draw some of these contractions as dotted lines because they are $T$'s, and others as thick 
lines to show that they are $\tilde{\omega}$'s.  In Fig.~\ref{f3}, $C^{i_{\rho_{10}} j_{\rho_{10}}}$, 
$C^{i_{\rho_{20}} j_{\rho_{20}}}$ and $C^{i_{\rho_{30}} j_{\rho_{30}}}$ are replaced with 
$\tilde{\omega}^{i_{\rho_{10}} j_{\rho_{10}}}$, $\tilde{\omega}^{i_{\rho_{20}} j_{\rho_{20}}}$ and 
$\tilde{\omega}^{i_{\rho_{30}} j_{\rho_{30}}}$, respectively, whereas $C^{i_{\rho_{11}} j_{\rho_{11}}}$, 
$C^{i_{\rho_{21}} j_{\rho_{21}}}$, $C^{i_{\rho_{22}} j_{\rho_{22}}}$ and $C^{i_{\rho_{23}} j_{\rho_{23}}}$ are 
replaced with $T^{i_{\rho_{11}} j_{\rho_{11}}}$, $T^{i_{\rho_{21}} j_{\rho_{21}}}$, 
$T^{i_{\rho_{22}} j_{\rho_{22}}}$ and $T^{i_{\rho_{23}} j_{\rho_{23}}}$, respectively.

Now notice another crucial observation.  This diagram can be reproduced by computing the Poisson bracket in 
${\cal W}$ first and mapping the resultant expression with $F$ later.  The Poisson bracket in ${\cal W}$ produces a 
product of three color-invariant observables.  The first one $L_1$ is characterized with the vertices 
$i_{\rho_{10}}$, $i_{\rho_{20}}$, $j_{\rho_{20}}$ and $j_{\rho_{10}}$; the second one $L_2$ with the vertices 
$i_{\rho_{20}}$, $i_{\rho_{30}}$, $j_{\rho_{30}}$ and $j_{\rho_{20}}$; and the third one $L_3$ with the vertices 
$i_{\rho_{30}}$, $i_{\rho_{10}}$, $j_{\rho_{10}}$ and $j_{\rho_{30}}$.  The mapping $F$ projects this product to 
${\cal N}$.  This is done, as usual, by splitting these 3 loops into smaller loops with dotted lines so that no two 
lines cross each other inside any of these 3 loops.  Notice that in Fig.~\ref{f3}, if we flip the dotted line 
$i_{\rho_{11(-1)}} i_{\rho_{11(1)}}$ to within $L_1$, and the dotted lines $i_{\rho_{20(-1)}} i_{\rho_{20(1)}}$
and $j_{\rho_{21(-1)}} j_{\rho_{21(1)}}$ to within $L_2$, no two dotted lines will cross each other.  Therefore, we 
obtain the same term.

Conversely, any term $F ( \{ f^I, f^J \}_W )$ can be seen as a term in $\{ F(f^I), F(f^J) \}_N$ by a simiar 
diagrammatic argument.  Thus $F$ is indeed a Poisson morphism.

This Poisson algebra may have other interesting mathematical properties.  We hope that we can use this Poisson 
algebra to solve gluon dynamics in the future.

We thank O. T. Turgut for discussions.  This work was supported in part by the U.S. Department of Energy under
grant DE-FG02-91ER40685.

\end{document}